\documentclass[journal=jacsat,manuscript=article]{achemso}

\usepackage[utf8]{inputenc}
\usepackage[T1]{fontenc}
\usepackage{graphicx}
\usepackage{amsmath,amssymb}
\usepackage{physics}
\usepackage{color}
\usepackage{multirow}
\usepackage{float}
\usepackage{braket}
\usepackage{adjustbox}
\usepackage{booktabs, tabularx}
\newcolumntype{C}{>{\centering\arraybackslash}X}
\usepackage[normalem]{ulem}
\usepackage{placeins}

\newcommand*{\vt}[1]{\textcolor{black}{ #1}}

\newcommand*{\jibin}[1]{\textcolor{black}{ #1}}

\author{Sanjoy Patra}
\affiliation{Solid State and Structural Chemistry Unit, Indian Institute of Science, Bangalore, Karnataka 560012, India}
\author{Jibin Sivanarayan}
\affiliation{School of Chemistry, Indian Institute of Science Education and Research, Thiruvananthapuram, Kerala 695551, India.}
\author{Vivek N. Bhat}
\affiliation{Solid State and Structural Chemistry Unit, Indian Institute of Science, Bangalore, Karnataka 560012, India}
\author{Philip D. Maret}
\affiliation{School of Chemistry, Indian Institute of Science Education and Research, Thiruvananthapuram, Kerala 695551, India.}
\author{Atandrita Bhattacharyya}
\affiliation{Solid State and Structural Chemistry Unit, Indian Institute of Science, Bangalore, Karnataka 560012, India}
\author{Sayan Ghosh}
\affiliation{Solid State and Structural Chemistry Unit, Indian Institute of Science, Bangalore, Karnataka 560012, India}
\author{Mahesh Hariharan}
\email{mahesh@iisertvm.ac.in}
\affiliation{School of Chemistry, Indian Institute of Science Education and Research, Thiruvananthapuram, Kerala 695551, India.}

\author{Vivek Tiwari}
\email{vivektiwari@iisc.ac.in}
\affiliation{Solid State and Structural Chemistry Unit, Indian Institute of Science, Bangalore, Karnataka 560012, India}

\title {State Localization and Selective Charge Filtering Near a Null Point}

\begin{document}

	\begin{abstract}
Null points in synthetically tunable molecular aggregates are predicted to generate flat energy bands analogous to those known in strongly correlated condensed-matter physics.  For chemistry, null points provide a powerful design principle for photovoltaic materials with selective charge filtering similar to photosynthesis. However, null points have never been experimentally verified because their defining prediction -- state localization with selective electron or hole transfer -- has remained unobserved. Here, using a donor–acceptor dyad as a minimal model, we provide the first experimental \jibin{observation} of a null point. Impulsive pump–probe measurements reveal charge separation through a near-instantaneously generated locally excited–charge transfer (LE–CT) intermediate that emerges upon solvent stabilization of CT states. Polarization anisotropy directly reveals state localization and selective charge-filtering, spanning balanced electron–hole transfer to selective hole filtering consistent with synthetic design. A generalized vibronic theory of null points explains these observations and identifies the ideal synthetic parameters for achieving null points which are protected from the vibrational bath.
	\end{abstract}

	\maketitle
	
	\section{Introduction}
	Photosynthetic reaction center proteins\cite{Blankenship2002} are pigment-protein complexes which can separate charge across tens of angstroms to create a microsecond long charge-separated state with near-unity quantum efficiency. The mechanism for the fastest steps of this process is now understood\cite{Policht2022,Zigmantas2017} to proceed through admixtures of vibrational-electronic (vibronic) states. This efficient design has inspired synthetically tunable molecular aggregates where a similarly fast charge separation, transport and eventual stabilization into a charge-separated state with low recombination probability is desirable. Such molecular constructs, which achieve charge separation between identical chromophores are termed as symmetry-breaking charge separation\cite{SebastianACSEngLett2022,WasielewskiACR2020} (SB-CS) systems, and may find potential applications as the active layer in organic photovoltaics. \\
	
{$\pi$-stacked molecular aggregates form locally excited ($LE$) Frenkel exciton type states as well as charge-transfer ($CT$) states due to orbital overlaps. Interference between long-range Coulomb ($J_C$) and $CT$ couplings results in the rich landscape of non-Kasha\cite{Hestand2018,Hestand2017} molecular aggregates. Using a purely electronic description, Spano and co-workers have predicted that such interference can also lead to flat energy bands, termed as null points\cite{Spano2024}, thus opening the possibility of synthetically tunable molecular materials with strongly correlated excitons similar to layered materials with flat energy bands\cite{Macdonald2011} known in condensed matter physics. For photochemistry, Spano has shown that null points are exceptionally susceptible\cite{Spano2024} to solvent induced energetic fluctuations, leading to SB-CS with {selective} hole or electron transfer. Charge filtering and simultaneous suppression, through destructive interference between \vt{orbitals}, of undesirable photophysical pathways which can arise in strongly coupled aggregates makes null points quite a versatile design in the broad context of fine tuning molecular photophysics towards desirable photoproducts. For example, selective charge filtering property with suppressed charge recombination channels finds promise as electron and hole transporting layers in photovoltaic applications. However, coupling to the vibrational bath, particularly at room temperature, can readily disrupt the intended functionality of null points, and remains an open question.} \\
	
The electronic description of Spano and co-workers also predicts\cite{Hestand2018} null excitons where, weak $LE-CT$ mixing through electron and hole overlap integrals, $t_e$ and $t_h$, respectively, can be treated perturbatively such that $J_{CT}$ and $J_C$ interfere destructively. Spectroscopically, this phenomenon should result in monomer-like absorption spectrum of the aggregate. Indeed, several recent experimental demonstrations of null excitons \cite{Lin2022,Lijina2023,Kaufmann2018,Hong2020,Sebastian2021,Lijina2020a,Sebastian2018}, based on approximate resemblance of the linear absorption spectrum to that of the monomer, have generated significant interest in the particular context of SB-CS.  \\

\jibin{Null points are a more general concept and should be distinguished from null excitons.} The presence of null excitons does not imply the existence of null points. \jibin{The latter arise from band degeneracies and have yet to be experimentally verified.} As elucidated by Spano,\cite{Spano2024} the former arises only in the perturbative limit of $CT$ coupling, whereas the latter constitutes a broader regime that enables selective charge filtering through state localization. To date, experimental evidence for null excitons relies primarily on the approximate similarity of linear absorption spectra to that of the monomer. In contrast, state localization with selective charge filtering—the key predicted signature of a null point—has not been experimentally realized. Verifying this prediction, and understanding the dynamical behavior of excitons in a vibrational bath near a null point, is the focus of this work.\\

The existence of null points and the resulting charge asymmetry is expected to be highly sensitive\cite{Spano2024} to the perturbations to $CT$ energies, fluctuations in the solvation environment, and the molecular geometry changes in going from a crystal environment, where the molecular geometry is characterized, to a solvated environment. Furthermore, vibronic couplings, so far not considered in the electronic description of null points, can interfere\cite{Makri2022} and delocalize\cite{Sahu2020,Thomas2026} electronic states. Whether selective charge filtering at null points intended by synthetic design even translates into the corresponding quantum dynamics is a question that remains unaddressed. Mechanistic questions such as the effect of linear vibronic couplings on the predicted orbital interference, and whether linear optical spectra adequately report on null excitons are all hard to decipher based on spectroscopic observables such as ratio of Franck-Condon (FC) intensities in the linear spectra, fluorescence lifetimes and non-impulsive PP experiments. \\

$D-A$ systems, in combination with impulsive optical probes and polarization-control, provide a model template for establishing mechanistic connections between the intended synthetic design and the dynamical behavior of excitons near null points. Deviating from all previous synthetic designs which rely on the destructive interference between finite $J_C$ and $J_{CT}$ couplings with balanced or comparable $t_e$ and $t_h$ couplings, we recently reported\cite{Sebastian2021} a Greek-cross architecture \vt{perylenediimide} (PDI) donor-acceptor ($D-A$) dyad where $J_C$ is suppressed by relative geometry and $J_{CT}$ is suppressed because of \jibin{strongly imbalanced} electron and hole overlap integrals, that is, $t_h \gg t_e$. The photophysics in polar solvents showed SB-CS in as fast as 627 fs with unprecedented charge separation to recombination rate ratio of 2947 inferred through a combination of linear spectra, fluorescence lifetime and non-impulsive pump-probe (PP) measurements. \jibin{Following Spano's terminology\cite{Spano2024} for lopsided electron and hole transfer integrals, we refer here to the more specific parameter regime of suppressed dipolar couplings together with strongly imbalanced $t_e$ and $t_h$ orbital overlaps as the lopsided regime. As our analysis will show, suppressed dipolar couplings are essential for protecting state localization near the null point against vibronic delocalization couplings.} \\


Recent impulsive PP experiments\cite{Lin2022, Hong2022} on \vt{PDI} based null exciton dimers and trimers report ultrafast charge separation through coherently excited intermediates with $CT$ character. However, as noted earlier, null points in SB-CS systems have never been experimentally realized. In this context, polarization control has provided unprecedented insights into ultrafast vibronic phenomenon such as in tracking electronic reorientation through a Jahn-Teller conical intersection\cite{kak2014,Farrow2008}. Here we present {sub-10 fs temporal resolution}, polarization-controlled, two-color PP experiments to provide the first experimental verification of null points in the Greek-cross PDI dyads. In polar solvents, tetrahydrofuran (THF) and acetonitrile (ACN), we find evidence for a near-instantaneously photoexcited $LE-CT$ intermediate state that subsequently evolves to a relaxed $CT$ state. Polarization anisotropy directly tracks electronic reorientation during charge separation and confirms the predicted state localization and charge asymmetry expected near null points. We find that selective charge filtering behavior is strongly influenced by the solvent polarity -- THF shows no such selectivity, whereas ACN, with significantly more $LE-CT$ mixing, shows selective hole filtering as intended by synthetic design. A generalized vibronic description of null points confirm the presence of mixed $LE-CT$ states and predicts a competing effect between electronic couplings conferred by synthetic design and linear vibronic couplings due to the intramolecular vibrational bath -- the former causes desired electronic localization leading to selective charge filtering whereas the latter can delocalize electronic states and prevent the process. Our model predicts that the dominating influence of the solvation environment and the competing effect from linear vibronic couplings can be simultaneously minimized in the lopsided regime -- significant imbalance between HOMO-HOMO versus LUMO-LUMO interference, and suppressed Coulomb interactions between the sites -- as is the case for the Greek-cross PDI dyad. Our work  provides the first experimental validation of null points and guides synthetic design towards null points that are protected from the vibrational bath.\\

\subsection*{Linear optical spectra show solvent polarity dependent deviations from the null exciton}

\begin{figure}[h!]
	\centering
	\includegraphics[width=5.5 in]{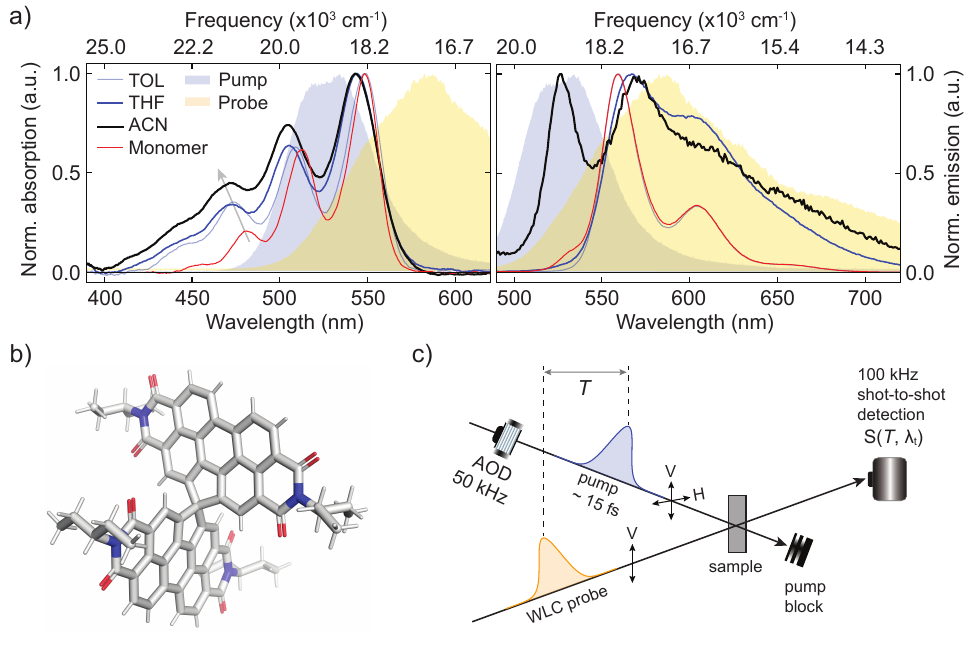}
	\caption{\footnotesize
		\textbf{Linear spectra of the Greek-cross spiro-conjugated PDI dimer (SpPDI2) and two-color polarization-controlled impulsive pump-probe (PP) setup.} (a) The absorption (left) and emission (right) spectrum of SpPDI2 in toluene (TOL), tetrahydrofuran (THF), and acetonitrile (ACN). The 24 nm red-shifted spectrum of the PDI monomer in TOL is also overlaid. The white light continuum (WLC) generated pump spectrum is blue shaded and centered at 525~nm. The WLC probe spectrum is gray shaded. (b) The Greek-cross (+) geometry of the SpPDI2 dimer obtained from the X-ray crystallographic structure\cite{Sebastian2021} is also shown for reference. (c) Schematic of the two-color PP (2C-PP) 100 kHz spectrometer. Polarization control on the pump and probe lines is denoted by vertical and horizontal arrows. $T$ denotes the PP delay time. Full optical layout is shown in \vt{Figure S1a}. 
	}
	\label{fig:fig1}
\end{figure}

\FloatBarrier

\vt{Figure~\ref{fig:fig1}a} shows the solvent polarity dependent linear spectra of the SpPDI2 dimer. Solvent polarity increases from lower in TOL (dielectric constant $\varepsilon = 2.38$) to THF ($\varepsilon = 7.58$) to highest in ACN ($\varepsilon = 37$). The fluorescence quantum yield\cite{Sebastian2021} reduces with increased polarity -- from \vt{$>$90\% in TOL to 24\% in THF and <1\% in ACN}.  Our previous work\cite{Sebastian2021} reported a mutual transition dipole angle of $\theta_{AB} = $ 87.4$^o$ obtained through single-crystal X-ray structure. The determined Greek-cross geometry is shown in Figure~\ref{fig:fig1}b. Dipole angle dependent electronic structure calculations, described briefly in \vt{Section S3}, show suppressed Coulomb coupling due to relative geometry despite only \vt{$\sim$3-4~\r{A}} separation between the PDI chromophores, and  constructive and destructive interference between HOMO-HOMO and LUMO-LUMO orbitals, respectively. This results in the lopsided regime, that is, minimal $J_C$ and $t_h \gg t_e$. This architecture also satisfies the electronic criterion for lower-exciton band null point, $J_c t_e t_h > 0$ outlined by Spano \cite{Spano2024}. Note that this is a purely electronic criterion and is generalized in this work to account for the differences in stabilization energies, quenching of electronic couplings and vibrational-electronic couplings that can arise for vibronic excitons. \\

When the red-shifted PDI monomer spectrum in TOL is overlaid with the SpPDI2 dimer in TOL, close agreement between the $I_{00}/I_{01}$ vibronic band intensities is seen, as previously reported\cite{Sebastian2021} -- \vt{1.74 for monomer versus 1.60 for the dimer}. Assuming a large energy gap between the $CT$ and $LE$ states that justifies a second-order perturbative treatment\cite{Hestand2018} of $LE-CT$ couplings, destructive interference between $J_C$ and $J_{CT}$, can indeed lead to a null exciton with monomer like peak intensities. All studies so far confirm null exciton formation through peak intensities in the linear optical spectra. {However, subtle differences in peak intensities, such as those seen in the $I_{02}$ and higher bands, hold vital clues regarding the sensitivity of null points to solvent polarity dependent variations in $LE-CT$ mixing and the consequent changes in state localization near null points.} With increase in polarity, deviations from the monomer absorption intensities, most notably under the higher vibronic progressions are increasingly evident. The corresponding emission spectrum (Figure~\ref{fig:fig1}, right) shows strong deviations from the monomer spectrum suggesting that relaxation pathways away from the Franck-Condon (FC) geometry are significantly modified even if $J_C$ and $J_{CT}$ destructively interfere. The lower lying emission peak in ACN was also confirmed by conducting the emission measurements at 400 nM concentrations where the intermolecular distance was greater than the F\"{o}rster critical radius\cite{Forster1959} of 17 \r{A} for energy transfer and minimizes fluorescence reabsorption effects. This is shown in \vt{Figure S1}. Figure~\ref{fig:fig1}c shows the polarization-controlled, 100 kHz shot-to-shot detected, two-color PP (2C-PP) setup where pump and probe white-light spectra are generated by focusing a $\sim$1 $\mu$m 1040 nm fundamental light into respective YAG crystals. The white-light spectra are overlaid with the absorption and emission spectra in Figure~\ref{fig:fig1}a. Further details of the method are presented in \vt{Section S1} and described in our previous works\cite{Bhat2023,Thomas2023}. The two-color nature of the experiment implies dominant excitation of only the main absorption band, while the probe spectrum is centered on the emission lineshape to dominantly probe the excited state dynamics. In a two-dimensional electronic spectroscopy experiment, this corresponds to probing the lower cross-peak region. The pump pulse duration was measured to be \vt{$\sim$ 15 fs} obtained by maximizing two-photon intensity autocorrelation (\vt{Figure S1}). The instrument response function obtained from a global fit across the $>$200 nm probe bandwidth is \vt{$35$ fs} (\vt{Figure S1}).  \\
\subsection*{Solvent polarity dependent, near-instantaneous generation of a $[LE+CT]$ intermediate}

\begin{figure}[h!]
	\centering
	\includegraphics[width=5.5 in]{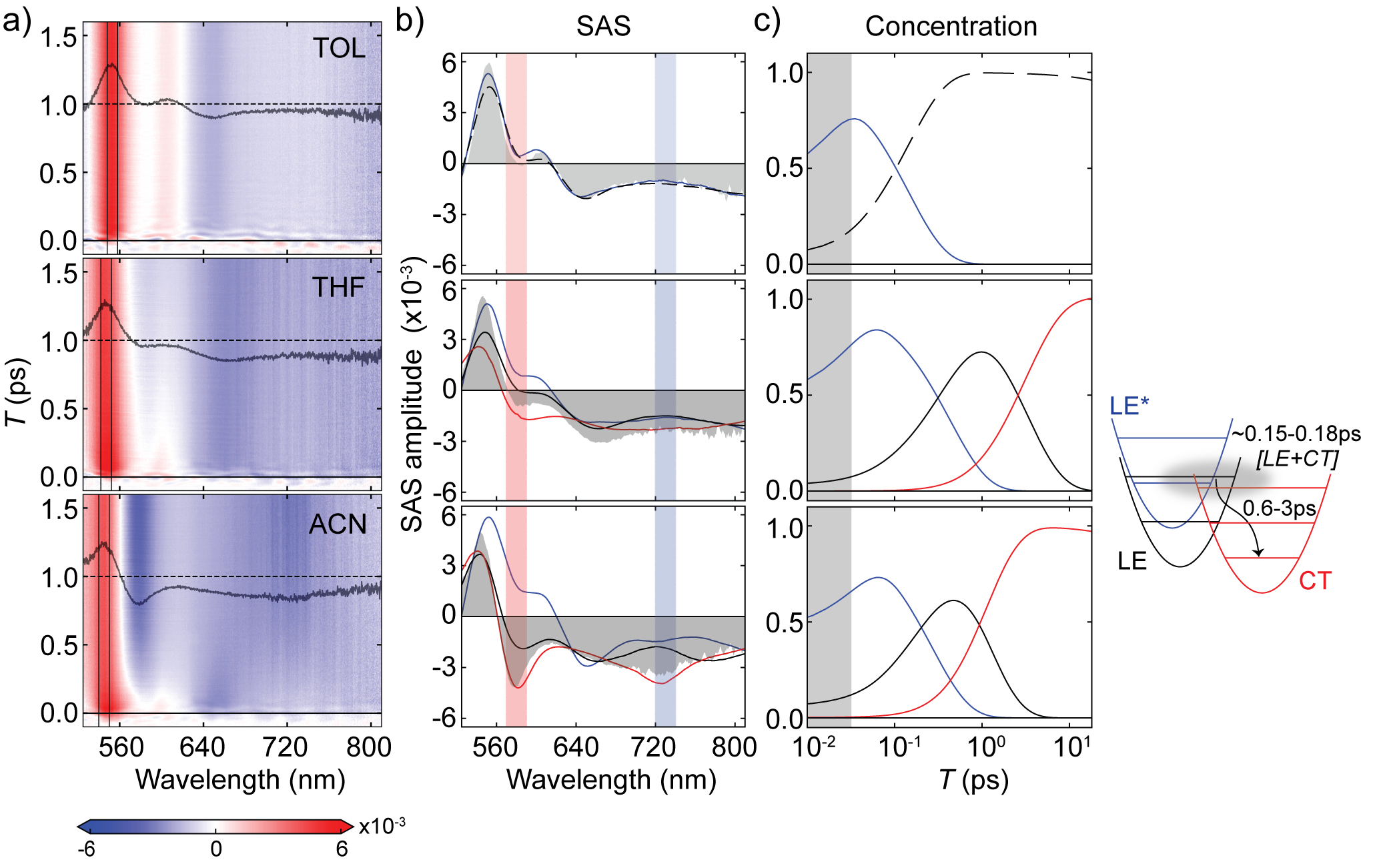}
	\caption{\footnotesize
		\textbf{Solvent-polarity dependent near-instantaneous generation of $[LE + CT]$ intermediate.} (a) Early time magic angle PP spectra for TOL (top), THF (middle) and ACN (bottom). The black trace in the contour plot represents the PP spectrum at $T = $1 ps. The full PP spectra and time constants obtained after global fitting are shown in \vt{Figure S2}. (b) Species associated spectra (SAS) derived from the global fit. The cation and anion bands are denoted by red and blue bands respectively. The gray shaded area represents the PP spectrum at $T = $1 ps. (c) The corresponding concentrations. The gray band at {35 fs} marks the 1-99 \% rise of the instrument response. The target model for the SAS analysis is shown on the right. The intermediate corresponds to a near-instantaneously photoexcited $[LE+CT]$ intermediate with a timescale of $\sim${150 fs and $\sim$180 fs} for THF and ACN respectively. The relaxation of this intermediate to the $CT$ state occurs on the timescale of 0.62 ps and 2.97 ps for THF and ACN respectively. 
	}
	\label{fig:fig2}
\end{figure}

\vt{Figure~\ref{fig:fig2}a} shows the early PP waiting time ($T$) PP spectra for the three solvents. A strong positive band resulting from the overlapping ground state bleach (GSB) and stimulated emission (SE) signals is seen for all cases. However the SE band shoulder persistent in TOL at 605 nm vanishes within $\sim$250 fs for both THF and ACN cases, and leads to the formation of cation and anion bands near 580 nm and 730 nm, marked by red and blue vertical lines, respectively. The loss of SE signal is consistent with the reduced fluorescence quantum yield with solvent polarity. Our previous report\cite{Sebastian2021} with slow instrument response confirmed the presence of cation and anion bands beyond 1 ps. These were found to be consistent with the PDI monomer cation and anion bands (see Figure S18 of ref.\cite{Sebastian2021}). With the impulsive 2C-PP experiments, a global fit and target analysis of the PP dataset in \vt{Figure~\ref{fig:fig2}b} shows species associated spectra (SAS) of an intermediate species with mixed $LE$ and $CT$ spectral features. The corresponding decay associated spectra (DAS) are shown in \vt{Figure~S2}. Similar to recent reports\cite{Lin2022, Hong2022} of coherently photoexcited intermediates in SB-CS, we assign this intermediate to a near-instantaneously generated mixed $[LE + CT]$ intermediate. The near-instantaneous generation is confirmed by the corresponding concentration profiles in \vt{Figure~\ref{fig:fig2}c} which show that \vt{37 \%} of the intermediate concentration is already present at the singlet maxima. \vt{Figure S5 overlays the PP spectra for early delay times to show the generation of the intermediate within the instrument response of $\sim$35 fs}. The intermediate concentration maximizes with time constants of \vt{$\sim$150 fs and $\sim$180 fs} for THF and ACN, respectively, and subsequently evolves to a relaxed $CT$ state on a 2.95 ps and {0.62 ps} timescale for THF and ACN, respectively. The above picture is schematically represented in the right panel of \vt{Figure~\ref{fig:fig2}c}. The cation and anion bands, marked in \vt{Figure~\ref{fig:fig2}b}, also show increased amplitude with increasing solvent polarity from THF to ACN. Note that, as shown in \vt{Figure~S2}, the same global model when applied to TOL leads to unphysical DAS with $<$1-2\% amplitude contributions. Consequently, the resulting SAS for the initially excited and the relaxed $S_1$ species in \vt{Figure~\ref{fig:fig2}b(top)} show nearly identical spectra with no indication of a $[LE+CT]$ intermediate in the case of TOL. \\

Similar coherently photoexcited $[LE+CT]$ intermediates in SB-CS have been recently \cite{Lin2022, Hong2022} reported to mediate SB-CS in PDI based systems. The accompanying vibrational quantum beats were reported to promote relaxation toward a relaxed $CT$ state away from the FC geometry.  We find an absence of vibrational quantum beats during the impulsive generation of the $[LE + CT]$ and its subsequent relaxation to the $CT$ state. \vt{Figures~S6-S9} show that in a related chiral PDI dimer\cite{HariharanMaret2023} \jibin{(CyPDI2)} in which no null points are expected based on Coulomb couplings, and electron and hole transfer integrals are not imbalanced (\vt{Section S3}), the same PP spectrometer and data analysis pipeline shows absence of $[LE+CT]$ intermediate. Compare \vt{Figure~S5} for SpPDI2 with \vt{Figure~S6} for the CyPDI2 dimer. However, the SB-CS process in the \jibin{CyPDI2} dimer is accompanied by prominent vibrational quantum beats in the range of \vt{$\sim$100-1600} cm$^{-1}$ (\vt{Figure~S7}). These results are summarized in \vt{Section S2}. We note that several previous works\cite{Joo2012,Petelenz2019,Tiwari2023JPCLett} show that spectator quantum beats, which have isotropic polarization signatures\cite{Bhattacharya2023,Sahu2025,Tiwari2026}, can simply accompany internal conversion or be excited by it, and do not report on a functional role for such nuclear motions. \


\FloatBarrier

\subsection*{Electronic anisotropy reports state localization and selective hole transfer}

Two strongly coupled and isoenergetic and electronically coupled $LE$ sites $|A\rangle$ and $|B\rangle$ form perfectly mixed excitons $(|A\rangle \pm |B\rangle)/\sqrt{2}$ where we denote\cite{Spano2024} the phase of the linear combination by a subscript on the wave vector $k$, as $\ket{k_0^{LE}}$ and $\ket{k_{\pi}^{LE}}$, respectively. In case of large spectral overlaps, such electronic mixing may not produce spectral changes during $T$. In such situations, polarization anisotropy can serve as a sensitive probe of electronic dynamics. Solvent polarity dependent $[LE+CT]$ intermediate formation and the expected null point behavior in this system prompted us to use polarization anisotropy to directly track the transition dipole reorientation that accompanies these steps. During internal conversion between the $LE$ and $CT$ states, the expected change in the electronic anisotropy\cite{Jonas1996} is $r(\theta_{ij}) = \frac{(3 \cos^2(\theta_{ij}) - 1)}{5}$, where $i, j$ are the transition dipole directions corresponding to the initial and final probed species. For instance, for orthogonally polarized $i,j$ transitions, electronic anisotropy of approximately -0.2 has been recently reported\cite{Vauthey2025} by Vauthey and co-workers in the context of $CT$ $D-A$ dyad. In case of the SB-CS dimer studied here, species $i$ is the initially excited $LE$ state while $j$ could be the intermediate or the relaxed $CT$ species formed later. Note that a simple consideration as above does not account for impulsive excitation, shared ground state electronic correlations between the sites, and $LE-CT$ couplings as expected in the SpPDI2 dimer. Later in this section, we account for these effects to formulate a generalized anisotropy model starting from a $CT$ dimer Hamiltonian and simulate the resulting electronic anisotropy near null points.

\begin{figure}[h!]
	\centering
	\includegraphics[width=5.5 in]{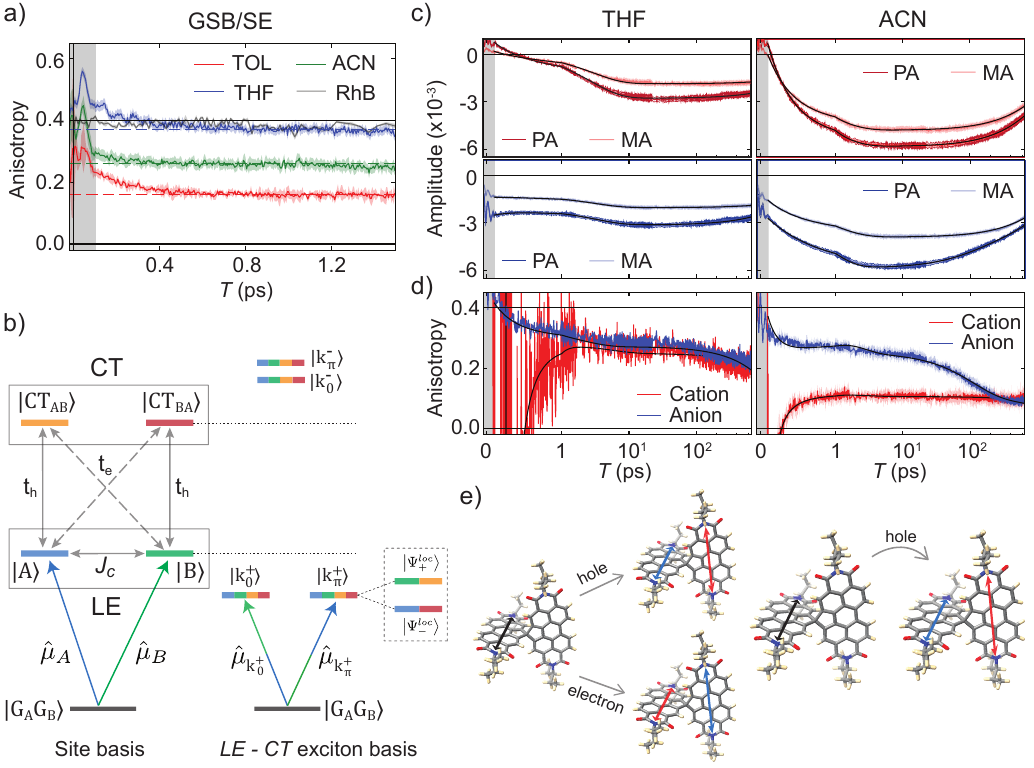}
	\caption{\footnotesize
		\textbf{Electronic anisotropy reports state localization and selective hole transfer expected near null points.} Electronic anisotropy in the (a) GSB/SE band for solvents of varying polarity. The corresponding PP transients are shown in \vt{Figures~S2 and S4}. The reference electronic anisotropy of 0.4 for a single $S_0\rightarrow S_1$ transition dipole is marked by a horizontal line and experimentally measured to be 0.381 $\pm$ 0.086 in the $T$ = 100-200 fs window for Rhodamine B (RhB) in ethanol (\vt{Figure~S3}). The other horizontal lines denote the settling value of the anisotropy of SpPDI2 in different solvents which is determined by averaging the \vt{1--2 ps} time window (\vt{Table S2}). The error bar is obtained by averaging 3 trials and shown as $\pm \sigma/\sqrt{N}$ band on the average. The anisotropy in the first 100 fs region, where the coherent artifacts persist in the corresponding PP transients (\vt{Figure~\ref{fig:fig2}a}), is grayed out. (b) A purely electronic $CT$ dimer model\cite{Spano2024} in the diabatic site basis \vt{qualitatively} explains the observed anisotropy trends as arising from state localization and selective hole filtering near the null point. $G_AG_B$ denotes the initially shared ground state between the sites. $A,B$ denote the collective $LE$ states where only one of the sites is electronically excited. The $LE$ states are mutually coupled through long-range dipole couplings $J_C$.  The $CT$ states couple to the $LE$ states through electron and hole overlap integrals, $t_e$ and $t_h$, respectively. The dashed box shows that starting from a $LE-CT$ (delocalized) exciton basis, the resulting states vectors show state localization\cite{Spano2024} and selective hole transfer at the null point. The electronic character is represented by color and the \vt{full analytic derivation is presented in Section S4.} (c) PP transients in the cation (top) and anion (bottom) bands for the case of THF and ACN. (d) The corresponding electronic anisotropy. The corresponding $PA$ and $MA$ fits are shown in thin black lines and the anisotropy fit is derived from these fit functions (\vt{Table S2}). (e) Illustration of cation (red) and anion (blue) transition dipole alignments\cite{Vauthey2011} involved in non-selective (left) versus selective (right) hole transfer.
	}
	\label{fig:fig3}
\end{figure}
\FloatBarrier

 \vt{Figure~\ref{fig:fig3}} plots the electronic anisotropy derived from the PP data in \vt{Figure~\ref{fig:fig2}}. The GSB/SE band and the cation and anion bands are marked in \vt{Figure~\ref{fig:fig2}}.  \jibin{Electronic polarization anisotropy can be derived\cite{Jonas1996} from PP experiments conducted with any two of the parallel ($PA$), perpendicular ($PE$) and magic angle ($MA$) relative pump and probe polarizations}. In our case, the anisotropy is derived from the $PA$ and $MA$ polarization sequence as $r(T) = (S_{PA} - S_{MA})/2S_{MA}$, where $S$ is the PP signal in any given spectral band. The anisotropy plots shown in \vt{Figure~\ref{fig:fig3}a} are averaged over three trials. The corresponding full PP transients for the GSB/SE band are plotted in \vt{Figure S2 and S4} and the \vt{settling anisotropy in the 1-2 ps time window is summarized in Table S2}. Using the same polarization optics, the reference anisotropy of 0.381 $\pm$ 0.086 between $T$ = 100-200 fs obtained for Rhodamine B (RhB) in ethanol, close to 0.4 expected from an isolated transition dipole, is shown in \vt{Figure~S3}. Full $T$ range anisotropy of RhB is shown in \vt{Figure S3} with anisotropy of \vt{0.004 $\pm$ 0.001} at $T$ = 600 ps. In case of SpPDI2 in TOL, where no $[LE+CT]$ intermediate was observed in Figure~\ref{fig:fig2}a, the early $T$ anisotropy in the GSB/SE band undergoes an initial drop that is attributable to excited state electronic relaxation and settles at 0.160 $\pm$ 0.014 by $\sim$600 fs. Jonas and co-workers have shown\cite{Qian2003} that the GSB electronic anisotropy ($r_{GSB}$) of excitonic dimers and square symmetric molecules with orthogonally polarized transition dipoles is 0.1 with isotropic signal strength of $S^{GSB}_{MA} = 2S^{SE}_{MA}$, because shared ground state implies that bleach on one molecules also affects the other. Similarly, $r_{SE}$ is expected to be 0.4 after all electronic coherences are dephased and assuming no further excited state electronic reorientation. The resulting total expected anisotropy in the GSB/SE band, $r_{tot} = (S^{GSB}_{MA}r_{GSB} + S^{SE}_{MA}r_{SE})/(S^{GSB}_{MA} + S^{SE}_{MA}) $ becomes 0.2 for a purely excitonic dimer in reasonable agreement to that seen for TOL. Contamination from the ESA signal overlapping with the GSB/SE band is likely to be minor contributing factor in the reduced anisotropy. In order to estimate whether oscillator strength borrowed by energetically higher lying $CT$ states through $LE-CT$ mixing affects the expected anisotropy, in \vt{Section S5} we extend the purely excitonic dimer description to a $CT$ dimer and analytically show that the observed anisotropy is insensitive to $LE-CT$ mixing. Therefore, the purely excitonic description qualitatively explains the reduced GSB/SE anisotropy for SpPDI2 in TOL as dominantly arising from a shared ground state. \\
 
\begin{figure}[h!]
 	\centering
 	\includegraphics[width=5.5 in]{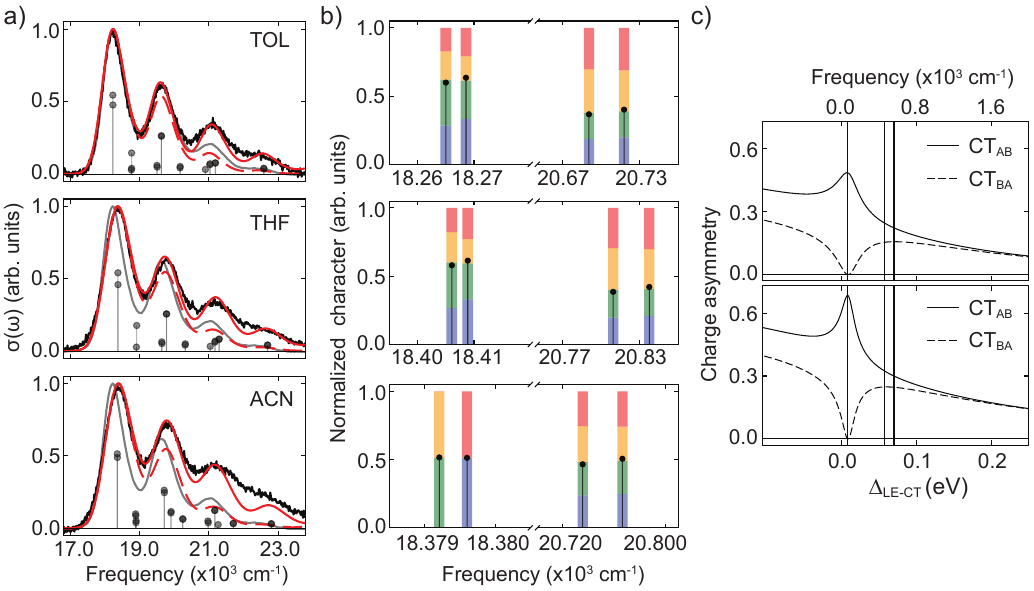}
 	\caption{\footnotesize
 		{Vibronic $CT$ dimer model connects solvent polarity dependent anisotropic charge transfer to null points.} (a) Simulated absorption spectrum for TOL (top), THF (middle) and ACN (bottom). Solid lines show the experimental spectrum of the dimer (black) and the 24 nm red-shifted monomer (gray). The \vt{dashed red curve} is the absorption spectrum simulated without any $LE-CT$ mixing. The \vt{red} curve corresponds to the optimized $\Delta_{LE-CT} = E_{CT} - E_{LE}$ energy gap which captures the increased absorption intensities under the higher vibronic shoulders. The corresponding stick spectra, without the Brownian oscillator lineshapes, are also shown for this case. (b) The corresponding purely electronic oscillator strengths along with state characters. The state characters for ACN are the same as in \vt{Figure~\ref{fig:fig3}b.} (c) Charge asymmetry and null points in the purely electronic (top) and the vibronic (bottom) dimer models corresponding to the case of ACN. Energetic fluctuations at the null point induced by the solvent are fixed at {$\delta E$ = 10$^{-4}$ eV}. The \vt{three vertical lines} correspond to the $\Delta_{LE-CT}$ energy gap that describes the optimized absorption intensities in panel a. The parameters are described in \vt{Table~S4}.
 	}
 	\label{fig:fig4}
 \end{figure}

 An immediately interesting aspect of this result is that the SpPDI2 dimer, even in the case of TOL where $I_{00}$ and $I_{01}$relative intensities are similar to those in the monomer (\vt{Figure~\ref{fig:fig1}}) and no $[LE+CT]$ intermediate is observed, is not a null exciton with monomer like optical properties -- the observed anisotropy reveals a shared ground state and significant electronic delocalization between the sites (Figure~\ref{fig:fig4}a). To the best of our knowledge, so far the existence of a null exciton has been based on comparisons of linear absorption intensities in the $I_{00}$ and $I_{01}$ bands. Our results for TOL -- electronic anisotropy and simulated linear vibronic spectra in Figure~\ref{fig:fig4}a -- demonstrate that deviations from monomer vibronic intensities in the $I_{02}$ and higher bands and the electronic anisotropy subtly report on non-negligible effects such as $LE-CT$ mixing and electronic delocalization between the sites, respectively, which are precluded under the assumption of perturbative $CT$ couplings. \\

Compared to TOL, the presence of an $[LE+CT]$ intermediate in THF and ACN cases suggests stabilization of $CT$ states with increasing solvent polarity. Curiously, the settling GSB/SE anisotropy rises to {0.370 $\pm$ 0.014} in case of THF. The same rising trend in GSB/SE anisotropy is also seen for the case of ACN which shows even greater amplitude of the $LE-CT$ intermediate (\vt{Figure~\ref{fig:fig2}b,c}) and settles at {0.258 $\pm$ 0.017}. Spano has illustrated\cite{Spano2024} how fluctuations in the solvation environment lead to state localization near null points due to mixing between the excitons of different phase. For instance, as illustrated in \vt{Figure~\ref{fig:fig3}b} and \vt{analytically shown in Section S4 for a model with $t_h \gg t_e$} as for the SpPDI2 dimer, transforming a $CT$ dimer model to a $LE-CT$ exciton basis shows the mixing between the $LE$ and $CT$ excitons. The  resulting $LE-CT$ excitons in the lower energy band are degenerate and denoted as $\ket{k_{0}^{+}}$ and $\ket{k_{\pi}^{+}}$, where $+$ denotes the phase of the linear combination. Solvent induced energetic fluctuations mixes them to result in a localized state with only $\ket{B}$ optically bright character with no electronic delocalization between the $A$ and $B$ sites (see \vt{Section S4.1} for full derivation). This is illustrated in \vt{Figure~\ref{fig:fig3}b} where the color tracks electronic character. In such a localized state with no shared ground state electronic correlations between the two sites, the resulting anisotropy will increase towards the isolated transition dipole limit of 0.4. This idea is illustrated in \vt{Figure S15} which calculates the electronic anisotropy at the exact null point with state localization versus away from it where the delocalized sites $A$ and $B$ share a common ground state. Therefore, solvent fluctuation driven state localization near null points explains the increased anisotropy in case of THF and ACN. The SE signal contribution, with its anisotropy of 0.4, is only expected in case of THF -- quantum yield $\phi$ = 24\% in THF versus <1\% in case of ACN -- and qualitatively explains the higher GSB/SE anisotropy in case of THF versus ACN. Note that our first guess for increased $r_{tot}$ in THF and ACN compared to TOL was the increased $LE-CT$ mixing suggested by simulations of linear vibronic intensities (Figure~\ref{fig:fig4}a). This expectation\cite{Jonas2011} is valid for excitonic dimers where borrowing of oscillator strength by higher lying states not covered by the laser spectrum, increases $r_{GSB}$ towards 0.4, the isolated transition dipole limit. However, our calculations in \vt{Section S5}, both analytic and numerical, show that $r_{GSB}$ for the lower exciton band of the $CT$ dimer remains at $\sim$0.1 despite $\sim$50\% $LE-CT$ mixing. \vt{Eqn.~S23 and Figure~S14 in Section~S5} show that the insensitivity of $r_{GSB}$ is specific to the lopsided parameter regime -- $t_h \gg t_e$ and suppressed $J_C$. Thus, our hypothesis of a lack of shared ground state due to state localization near null points, outlined in \vt{Figure~S15}, is most consistent with the observed anisotropy trends.\\




As illustrated in Figure~\ref{fig:fig3}b, the key verification of a null point is whether state localization, inferred here through polarization anisotropy in the GSB/SE band, is also accompanied by selective hole filtering given that the hole transfer integral $t_h$ is synthetically engineered to be significantly larger than the electron transfer integral $t_e$? \vt{Figure~\ref{fig:fig3}c} shows the $PA$ an $MA$ transients in the cation and anion bands for the case of THF and ACN solvents. The corresponding electronic anisotropy is shown in \vt{Figure~\ref{fig:fig3}d}. For the case of THF, the anion band shows larger anisotropy initially. However, this anisotropy is only maintained until $\sim$1 ps, beyond which both bands show overlaying anisotropy as the $[LE+CT]$ intermediate relaxes to a $CT$ state. In contrast, the ACN case, which shows more prominent spectral signatures of the $[LE+CT]$ intermediate (\vt{Figure~\ref{fig:fig2}b}), behaves exactly as synthetically engineered -- the anion band anisotropy is significantly larger than the cation band with both merging on the timescales of molecular rotation. Following the argument\cite{Vauthey2011} of Vauthey and co-workers, \vt{Figure~\ref{fig:fig3}e} illustrates that the observed anisotropic charge transfer is a direct consequence of selective hole transfer, that leads to the localized state $\ket{B} - \ket{CT_{AB}}$ (\vt{Section S4} and \vt{Figure~\ref{fig:fig3}b}). A larger anisotropy is seen in the anion band because the initial $LE$ transition dipole on a given site, say $\hat{\mu}_B$, and the final transition dipole on the anionic site that contributes to the ESA transition, $\hat{\mu}_{B_- \rightarrow B_-^*}$, are parallel\cite{Gordon2003}. \\

We have also conducted experiments on a chiral CyPDI2 dimer\cite{HariharanMaret2023} (Section S2), where electronic structure calculations (\vt{Section S3 and Figure~S9}) do not predict null points. \jibin{The predicted absence of a null point in CyPDI2 is attributable to the lack of imbalance between $t_e$ and $t_h$ resulting from only partial HOMO-HOMO constructive and LUMO-LUMO destructive orbital interference. This essential difference between the SpPDI2 and CyPDI2 dimers is illustrated in {Figure~S9c,d}.} \jibin{Experimentally SB-CS is observed in CyPDI2 but without any evidence for a $[LE+CT]$ intermediate (Figure~S6). In contrast to SpPDI2, no signatures of state localization are observed in the GSB/SE anisotropy (Figure~S6e), and the cation and anion band anisotropies do not report any anisotropic charge transfer (\vt{Figure~S8b})}. Along similar lines, earlier work\cite{Vauthey2011} by Vauthey et al. on a perylene dyad without synthetically engineered null points also reported no intermediates or anisotropic charge transfer during SB-CS. \\

Interestingly our observations show that synthetic design for null points and the solvation environment, both strongly influence the directionality of charge transfer -- non-selective \jibin{on longer timescales} in THF to selective hole filtering, as intended by synthetic design, in ACN. To understand these observations better, next we simulate the solvent-dependent vibronic intensities for the SpPDI2 dimer. \\
 
 \FloatBarrier
 
 \subsection*{Vibronic description of null points connects solvent polarity dependent anisotropic charge transfer to linear spectra}
 
The electronic $CT$ dimer model starts from the diabatic site basis states -- $\ket{G_AG_B}$ for the shared ground state, $\ket{A}$ and $\ket{B}$ for the locally excited states, and $\ket{CT_{AB}}$ and $\ket{CT_{BA}}$ for the $CT$ states $\ket{A^+B^-}$ and $\ket{A^-B^+}$, respectively. The electronic Hamiltonian in the site basis, $\hat{H}_{elec}^{site}$ is shown in \vt{Eqn.~S1}. The general analytic derivation of the resulting electronic eigenstates for any $J_C$, $t_e$ and $t_h$ electronic couplings, including the lopsided parameter regime for the SpPDI2 dimer, is presented in \vt{Section S4}. The resulting localized states at null points, caused by an energy fluctuation $\delta E$ induced by the solvent, are illustrated in \vt{Figure~\ref{fig:fig3}b}. Extending this picture to a vibronic description, we include a high- and a low-frequency intramolecular vibrational mode explicitly in the system Hamiltonian to fully account for rich effects such as electronic delocalization caused by vibronic mixing\cite{Tiwari2013, Sahu2020}, interference of vibronic couplings along low- and high-frequency vibrations\cite{Bhattacharya2023,Hong2022,Lin2022} and their interplay with $CT$ couplings within a numerically exact basis set description without any N-particle approximation\cite{Sahu2020}. Linear vibronic couplings from the rest of the intramolecular vibrational bath are treated as \vt{Brownian oscillators}, and impart broadened lineshapes to the absorption sticks in Figure~\ref{fig:fig4}a. Calculations for electronic couplings -- $J_C$, $t_e$ and $t_h$ follow our earlier report\cite{Sebastian2021} on the SpPDI2 dimer and described in \vt{Section S3}. All the model parameters are summarized in \vt{Table S3 and S4}. It should be emphasized that only the $LE-CT$ energy gap, $\Delta_{LE-CT} = E_{CT} - E_{LE}$, is floated to account for the solvent polarity dependent stabilization of the $CT$ states. All other parameters derived from electronic structure calculations are kept fixed across the solvents. The resulting vibronic intensities are plotted in Figure~\ref{fig:fig4}a and compared with the dimer (black) and monomer (gray) absorption spectra.\\

As noted earlier in the discussion of Figure~\ref{fig:fig1}, the deviations in the $ I_{01}$ and higher vibronic shoulders compared to the monomer follow the solvent polarity. Simulations of vibronic intensities in \vt{Figure~\ref{fig:fig4}a} show that the observed trends in the absorption intensities are explained by the largest $LE-CT$ mixing in case of ACN due to stabilized $CT$ states in the polar solvent (compare the dashed versus solid red curves in Figure~\ref{fig:fig4}a). For THF and TOL, the $CT$ states are progressively less stabilized such that $LE-CT$ mixing is reduced and so are the corresponding vibronic intensities under $ I_{02}$ and higher energy shoulders. The corresponding electronic state characters are plotted in \vt{Figure~\ref{fig:fig4}b}. With increasing solvent polarity, and therefore increasing $LE-CT$ mixing, the color-coded electronic characters show maximum state localization for the case of the ACN solvent. The localized state characters plotted in \vt{Figure~\ref{fig:fig3}b} correspond to the electronic null point in \vt{Figure~\ref{fig:fig4}b}. \\

To further visualize\cite{Spano2024} state localization, which should result only in $CT_{AB}$ or $CT_{BA}$ character, \vt{Figure~\ref{fig:fig4}c(top)} plots the corresponding $CT$ character of the lowest exciton of the purely electronic model as a function of $\Delta_{LE-CT}$. Larger $LE-CT$ mixing through reduced $\Delta_{LE-CT}$ also ensures proximity to the null point and consequently maximum charge asymmetry and selectivity in hole transfer. The $\Delta_{LE-CT}$ values estimated by fitting the vibronic intensities for the three solvent cases are indicated by the vertical lines in Figure~\ref{fig:fig4}a. When $LE-CT$ mixing is less such as in the case of THF and TOL solvents, deviations in $> I_{02}$ vibronic intensities compared to the monomer are also lesser. At the same time, the system is farther away from the null point. Consequently, the charge asymmetry is lesser such that both electron and hole transfer become equally probable as reported by the equal cation and anion band anisotropies for the case of THF.  \\

Overall, by relating absorption intensity trends to increased $LE-CT$ mixing, the calculations in Figure~\ref{fig:fig4} elucidate the influence of solvent polarity on state localization and selective hole transfer near null points. We note that our theoretical model only {qualitatively} explains our experimental observations -- solvent polarity dependent vibronic shoulder intensities (Figure~\ref{fig:fig1}), and state localization with anisotropic charge transfer (Figure~\ref{fig:fig3}). \vt{This is so because the optimized fits} to the linear spectra are tolerant to $\pm$10-20\% variations in the estimated $\Delta_{LE-CT}$ gaps, and to $\sim10\times$ variations in the already suppressed $J_C$ Coulomb coupling (\vt{Figure~S13}). Such variations in the estimated parameters are well within the \vt{estimation error\cite{Matsika2013} from electronic structure calculations across different basis sets and in the presence of a solvent}.  \\

 \subsection*{Lopsided parameter regime protects null points from the vibrational bath}
The above observations naturally raise the question of how -- if at all -- electronic null points are modified by coupling to the vibrational bath. In \vt{Section~S4.2}, we develop a vibronic framework for describing null points that elucidates the role of linear vibronic coupling in either stabilizing or suppressing them. This analysis further identifies the parameter regimes in which null points remain protected against the vibrational bath.\\

The derivations in \vt{Section S4} generalize the vibronic exciton framework\cite{Tiwari2017,Tiwari2018,Bhattacharya2023} to $CT$ dimers and only the key points regarding the modification of null points due to vibronic couplings are briefly described below. Denoting the numerically exact vibronic basis states as $\lvert A \rangle \lvert 0_B \rangle \prod_{j\in\{H,L\}} \lvert \tilde{v}_j^A \rangle \lvert v_j^B \rangle$, where $H,L$ correspond to the high and low frequency intramolecular vibrational modes in the system Hamiltonian, and $\tilde{v}$ denotes the FC displaced vibrational basis states on the electronically excited site, $A$ in this case. As detailed in \vt{Section S4} and illustrated in Figure~\ref{fig:fig3}b, transformation from site basis to the $LE-CT$ exciton basis leads to eigenvectors $\lvert k_{\pi}^+ \rangle$ and $\lvert k_{0}^+ \rangle$ in the lower energy band. \vt{As shown in Eqn.~S7 and Eqn.~S10}, a solvent induced energetic fluctuation $\delta E$ couples the two states with coupling matrix element $\delta E \sin \theta_1 \sin \theta_2 \prod_{j\in\{H,L\}}\langle \tilde{v}_j^A \mid v_j^A \rangle\langle v_j^B \mid \tilde{v}_j^B \rangle$, where $\theta_1$ and $\theta_2$ are electronic mixing angles (\vt{Eqn.~S3}) between $LE$ and $CT$ states which depend only on $J_C$, $t_e$, $t_h$ and $\Delta_{LE-CT}$. The vibrational overlap factors, such as $\langle \tilde{v}_j^A \mid v_j^A \rangle$ between $\ket{A}$ and $\ket{0_A}$ partly quench the electronic coupling through $\delta E$. However, as seen from \vt{Eqn.~S15 and Table S5}, the FC displacements of the $U_{\lvert k_{\pi}^+ \rangle}$ and $U_{\lvert k_{0}^+ \rangle}$ diabatic exciton potentials are approximately equal in the lopsided regime. This is illustrated in \vt{Figure~S12} where no relative FC displacement is seen between them. This imposes a vibrational selection rule whereby state localization due to a fluctuation $\delta E$ occurs predominantly without any change in vibrational quanta. We will term this coupling due to $\delta E$ as `localization coupling'. \\

Similar vibrational overlap factors indeed partially quench all the electronic couplings to modify the null point. For example, $J_C$ is quenched by vibrational overlap factors $\prod_{j\in\{H,L\}} \langle\tilde\nu^A_j|\nu^A_j\rangle \,\langle\nu^B_j|\tilde\nu^B_j\rangle$ (\vt{Eqn.~S10}). Additionally, vibrational stabilization energy on each excitonic state also modifies the null point degeneracy as $E_{k_0^+} - E_{k_{\pi}^+} = \Delta \lambda_{k_0^+ - k_{\pi}^+}$ (\vt{Eqn.~S16}) where the right hand side is the difference between the total vibrational stabilization energy $\lambda$ associated with $U_{\lvert k_{\pi}^+ \rangle}$ and $U_{\lvert k_{0}^+ \rangle}$ diabatic exciton potentials. However, suppressed $J_C$ and the lack of relative FC displacements between the $U_{\lvert k_{\pi}^+ \rangle}$ and $U_{\lvert k_{0}^+ \rangle}$ potentials in the lopsided regime (\vt{Figure~S12}), ensure that the null point degeneracy condition is not affected. The minor effects from linear vibronic couplings are seen in Figure~\ref{fig:fig4}c (bottom) where for the vibronic $CT$ dimer model, the width of charge asymmetry $\eta$ (\vt{Eqn.~S19}) due to $\delta E$ fluctuation is only partly reduced, and to have the null point degeneracy at the same $\Delta_{LE-CT}$ as the purely electronic case (top), only a minor modification of $J_C$, from \vt{15.5 cm$^{-1}$ to 20.3 cm$^{-1}$}, is required. Overall, the influence of linear vibronic couplings on the null point, mediated by intramolecular vibrational motions, is strongly suppressed in the lopsided parameter regime. \\

\subsection*{Low-frequency vibrations can inhibit state localization at null points}

In the context of excitonic dimers, linear vibronic couplings are also well understood to promote electronic delocalization\cite{Womick2011, Tiwari2013} at vibronic resonances, where excitation of vibrational quanta can compensate for energetic mismatch between states. In terms of state mixing, imperfectly mixed electronic character is enhanced to near-perfect mixing at a vibronic resonance. For example, see the red curve in Figure~4 of ref.~\cite{Sahu2020} which quantifies electronic mixing by an inverse participation ratio. The effect of linear vibronic couplings is best seen when the vibrational part of the Hamiltonian, $\hat{H}_{LVC}$ in \vt{Eqn.~S9}, is written in terms of correlated and anti-correlated vibrational coordinates\cite{Moffit1960}, $\hat{q}_{j+}$ and $\hat{q}_{j-}$, respectively \vt{(Eqns.~S11-S12)}, where the latter is akin to the tuning coordinate in conical intersections\cite{JonasARPC2018}. In the $LE-CT$ exciton basis, $\hat{H}_{LVC}$ transforms to couple the $U_{\lvert k_{\pi}^+ \rangle}$ and $U_{\lvert k_{0}^+ \rangle}$ potentials through coupling elements $- {(\omega_j^2 S_j)}^{1/2}(\hat{q}_{j-}\cos \theta_1 \cos \theta_2)$ (\vt{Eqns.~S13-S14}), where $S_j$ is the Huang-Rhys factor along the $j^{th}$ intramolecular vibrational coordinate. \\

\begin{figure}[h!]
	\centering
	\includegraphics[width=5 in]{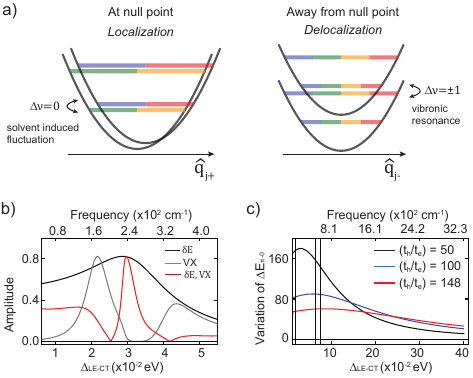}
	\caption{\footnotesize
		{Low-frequency vibrations can inhibit null points} (a) State localization due to solvent induced $\delta E$ energetic fluctuations versus electronic delocalization due to vibronic resonances along low-frequency energy gap tuning modes $\hat{q}_{L-}$. The former occurs dominantly without any change in vibrational quantum number. The latter coupling channels delocalize the imperfectly mixed non-degenerate electronic states by compensating the energetic mismatch between the two $LE-CT$ excitons by a low-frequency vibrational quanta. In the lopsided regime the $\Delta v= 0$ and $\Delta v = \pm1$ channels become exclusive. (b) Black curve plots the charge asymmetry versus $\Delta_{LE-CT}$ caused by a fluctuation $\delta E$, without any linear vibronic coupling in the system. State localization peaks at that $\Delta_{LE-CT}$ which leads to null point degeneracy, $\Delta E_{{\pi-0}} = |E_{k_0^+} - E_{k_{\pi}^+}| = 0$ and has a associated width. The gray curve, with $\delta E$ absent but including the linear vibronic coupling shows the oscillator strengths of the upper exciton $\lvert k_{\pi}^+ \rangle$ near a vibronic resonance with a \vt{36 cm$^{-1}$} low-frequency mode, denoted as VX. The energetic mismatch $\Delta E_{\pi-0}$ is compensated by a low-frequency vibrational quantum, leading to intensity borrowing and enhanced oscillator strength near resonance. The red curve plots charge asymmetry when both $\delta E$ and VX are present in the system. (c) Variation in $\Delta E_{\pi-0}$ with $\Delta_{LE-CT}$, $\frac{\partial \Delta E_{\pi-0}}{\partial \Delta_{LE-CT}}$ plotted as a function of $\Delta_{LE-CT}$ for fixed $J_C$ and $t_e$, same as in Figure~\ref{fig:fig4}, and different $(t_h/t_e)$ ratios. The highest ratio corresponds to the SpPDI2 dimer. Further details of the calculations are presented in \vt{Section S4.3}.}
	\label{fig:fig5}\end{figure}

As noted earlier, the diabatic exciton potentials $U_{\lvert k_{\pi}^+ \rangle}$ and $U_{\lvert k_{0}^+ \rangle}$ exhibit no relative FC displacement (\vt{Figure~S12}). Consequently, vibronic couplings along $\hat{q}_{j-}$ (\vt{Eqn.~S13}) involve transitions with $\Delta v = \pm 1$, whereas localization couplings are dominant for $\Delta v = 0$, i.e., in the absence of changes in vibrational quanta. A change in vibrational quanta along the tuning coordinate $\hat{q}_{j-}$ couples $U_{\lvert k_{\pi}^+ \rangle}$ and $U_{\lvert k_{0}^+ \rangle}$ diabatic exciton potentials away from a null point when they are not degenerate. This coupling delocalizes the electronic character and therefore has the opposite effect of $\delta E$ localization couplings. We term it as delocalization coupling. The competing nature of both couplings -- localization coupling due to solvent induced $\delta E$ fluctuations and delocalization coupling due to vibronic resonances -- are summarized in \vt{Figure~\ref{fig:fig5}a}. Note that as seen in \vt{Eqn.~S14}, the $\hat{q}_{j+}$ coordinate is akin to the seam space in conical intersections and does not participate in mixing the $LE-CT$ excitons. \\

It is important to emphasize that the exclusivity and selectivity of localization and delocalization coupling channels -- arising via $\Delta v = 0$ and $\Delta v = \pm 1$, respectively -- are direct consequences of the lopsided parameter regime. In this regime, the null point degeneracy is not only protected against quenching of electronic couplings and vibrational stabilization, but the competing delocalization arising from vibronic couplings is relegated to a distinct channel, separate from the desired $\delta E$ induced localization.\\

Just like $\delta E$ localization couplings which impart a width to a null point \vt{(Figure~\ref{fig:fig4}c}), resonant vibronic couplings also have a width proportional to $\sqrt{S_j}\cos \theta_1 \cos \theta_2$, that is, to the strength of vibronic coupling along a given coordinate and electronic mixing angles between the $LE$ and $CT$ excitons (\vt{Eqn.~S3}). Thus, for low-frequency vibrations and environmental modes\cite{Tiwari2017}, which can tune energy gaps between $A$ and $B$ sites, null point resonance which causes electronic localization and vibronic resonance which causes electronic delocalization can overlap for a range of $\Delta_{LE-CT}$ energy gaps. The latter can compete against localization couplings to suppress state localization and charge asymmetry near the null point. This effect is illustrated in \vt{Figure~\ref{fig:fig5}b} which plots charge asymmetry near a null point for a system with low-frequency vibrations. The low-frequency (\vt{$\sim$36 cm$^{-1}$}) intramolecular vibrational mode couples the lower-energy $LE-CT$ excitons away from the null point. Owing to the finite width of this resonance and the associated electronic delocalization, state localization and the resulting charge asymmetry becomes increasingly confined to the immediate vicinity of the null point. \\

\FloatBarrier

The analysis in Figure~\ref{fig:fig5}b suggests that low-frequency vibrations with large reorganization energies will compete against state localization electronic null points through vibronic resonances near null points. Large Huang-Rhys factors on low-frequency vibrations are not desirable because they ensure that such resonances are broad\cite{Bhattacharya2023} such that exact resonances do not matter and even multiple\cite{Tiwari2018} low-frequency near-resonant modes can collectively suppress localization. \\

The next question is whether there exist synthetic parameters, analogous to the lopsided parameter regime, that render the energy gap $\Delta E_{\pi-0}$ between lower-band excitons insensitive to solvent-polarity–induced stabilization of $CT$ states. Such a design can suppress competing delocalization couplings by lowering the cutoff beyond which low-frequency vibrational modes cannot achieve vibronic resonance with the near-zero $\Delta E_{\pi-0}$ at the null point. The ideal parameter regime must ensure minimal changes in $\Delta E_{\pi-0}$ with solvent induced variations in $\Delta_{LE-CT}$. Using the analytic expression in \vt{Eqn.~S5}, the variation of $\Delta E_{\pi-0}$ with $\Delta_{LE-CT}$, that is, $\frac{\partial \Delta E_{\pi-0}}{\partial \Delta_{LE-CT}}$ is plotted in Figure~\ref{fig:fig5}c for fixed values of $J_C$ and $t_e$ and varying $(t_h/t_e)$ ratio. For a fixed $J_C$, least variation in $\Delta E_{\pi-0}$ is seen when $t_h \gg t_e$. \vt{ Figure~S11} shows a similar plot as Figure~\ref{fig:fig5}c but with a fixed lopsided $(t_h/t_e)$ ratio and varying $J_C$. Again, variations in $\Delta E_{\pi-0}$ are minimal when dipolar couplings are suppressed as for the Greek-cross dimer studied here. These latter findings are consistent with those recently proposed by Spano\cite{Spano2024}. The analysis in Figure~\ref{fig:fig5} implies that given a variation in the $\Delta_{LE-CT}$ gap created by the solvent, the corresponding variations around the null point, that is, in $\Delta E_{\pi-0} \sim 0$ will be least. This implies two things -- 1.) The null point degeneracy in the lopsided regime is also the most robust to solvent polarity. And, 2.)  the null point is only affected by the lowest frequency vibrations. Dynamically it also means that a slower vibrational mode will be less effective in causing electronic delocalization such that electronic localization can prevail sooner than delocalization.

 Overall, our analysis shows that a combination of lopsided $t_h$ and $t_e$ couplings along with suppressed dipolar Coulomb couplings, as for the SpPDI2 dimer studied here, is most ideal to protect the null point -- first, by minimizing the impact of local variations in solvation environments on the null point degeneracy, and second by minimizing the impact of intramolecular vibrations on the desirable state localization by creating exclusive coupling channels for localization versus delocalization couplings.  \\

\section{Conclusions}

We provide the first experimental validation of a null point and predict the ideal synthetic regime to achieve the same. This is established through its hallmark signature—state localization with selective charge filtering. Polarization anisotropy is demonstrated as a sensitive probe of transition dipole reorientation, directly linking synthetic design to null point. We formulate a vibronic description of null points that captures the interplay of resonances and coupling interference, to enable direct connections with experimental observables. In particular, solvent-dependent vibronic intensities, state localization, and selective hole filtering are consistently explained by polarity-dependent $LE-CT$ mixing near the null point.\\

Our results also underscore the dominant role of the solvation environment while identifying the parameter regime that mitigates its influence. In particular, the lopsided parameter regime with suppressed dipolar couplings is predicted to optimally preserve null-point degeneracy and state localization by minimizing the competing effects of low-frequency vibrations and solvent polarity. Overall, this work establishes the key design criterion and optical signatures for future studies of engineered null points.

\section*{Methods}

\subsection*{Two-color pump-probe and anisotropy experiments}
{\footnotesize Two-color white-light PP measurements were performed as described previously\cite{Thomas2023,Bhat2023}. A 1040 nm output from a Yb:KGW amplifier was split equally into pump and probe arms, each carrying about 1 $\mu$J. The pump was modulated at 50 kHz, and pump/probe white-light continua were generated in \vt{4 mm and 8 mm} YAG crystals, respectively. Residual fundamental light was removed with shortpass filters, and dispersion in both arms was partially compensated with chirped mirrors. As shown in \vt{Figure~S1}, the pump pulse duration at the sample was $\sim$15 fs, and the spectrally resolved PP data yielded an instrument response function of $\sim$35 fs FWHM. Pump and probe were spatially overlapped and focused with an off-axis parabolic mirror to 70 $\mu$m 1/e$^2$ spot diameters in a 200 $\mu$m cuvette at a crossing angle of $\sim$7.5$^{\circ}$. Typical pulse energies at the sample were 2.80 nJ (pump) and 1.24 nJ (probe). The delay was controlled with a motorized translation stage. Measurements were acquired in parallel and magic-angle polarization configurations. The polarization settings were verified using the procedures outlined previously\cite{Tiwari2026} and tested on Rhodamine B as a reference sample (\vt{Figure~S3}).

For rapid-scan experiments, the population time delay was scanned from -0.5 ps to 660 ps in four constant-velocity intervals. Global analysis of the PP data was performed in GloTarAn\cite{Stokkum2012} and with custom Python scripts using multi-exponential kinetic models convolved with a Gaussian instrument response function. Data were chirp-corrected before fitting, and the DAS were extracted from the global fits. Target analysis was then used to test specific kinetic schemes and obtain SAS. Error bars were calculated from repeated parallel- and magic-angle measurements, and anisotropy uncertainties were propagated from the corresponding averaged transients.}

\subsection*{Sample and Electronic Structure Calculations}
{\footnotesize {Details of sample synthesis and characterization have been reported previously\cite{Sebastian2021,HariharanMaret2023}. SpPDI2 was dissolved in TOL, THF and ACN solvents to achieve the desired OD of 0.2-0.3 in 200$\mu$m cuvette. CyPDI2 exhibited very poor solubility in ACN rendering PP measurements in this solvent infeasible. All crystallographic computations were performed using the WINGX software suite. Three-dimensional molecular representations were generated using Mercury 3.10.1.4.} Quantum chemical calculations were performed using Gaussian 16. Geometry optimizations were carried out at the DFT level employing the $\omega$B97XD functional with the def2-SVP basis set in the gas phase unless otherwise specified. Vertical excitation energies and oscillator strengths were computed using time-dependent DFT (TD-DFT) at the same level of theory. Solvent-dependent TD-DFT calculations were conducted using the polarizable continuum model (PCM) with identical functional and basis set settings. Transition dipole moments for SpPDI2 were evaluated and visualized using Multiwfn 3.7.8 and VMD.} \\

\subsection*{Vibronic Exciton Model and Calculations}
{\footnotesize The vibronic description of null points formulated in \vt{Section~S4} extends the vibronic exciton framework of our previous works\cite{Tiwari2017,Tiwari2018,Bhattacharya2023} to $CT$ dimers with $LE$ and $CT$ diabatic site states. A high- and a low-frequency intramolecular vibrational mode on each PDI chromophore are treated explicitly in the system Hamiltonian, without invoking the one-particle approximation\cite{Sahu2020}, making the description numerically exact within the chosen basis. \vt{Each quantum oscillator has total 5 vibrational quanta which ensures convergence of $>I_{02}$ vibronic intensities}. The parameters are listed in \vt{Table~S4}. For simulations of absorption intensities, the rest of the vibrational bath is modeled as Brownian oscillators (\vt{Table~S3}) which account for optical decoherence and impart broadening to the oscillator strength stick spectra. The height of vibronic shoulders in the linear spectra was tolerant to 10-20\% changes in the $\Delta_{LE-CT}$ energy gap. Since $J_C$ is suppressed to begin with, 10$\times$ changes in $J_C$ do not significantly affect vibronic intensities. The FC displacement $d_{CT}$ on the $CT$ states was found to be the most sensitive parameter that affects shoulder vibronic intensities and was optimized for the best fit (\vt{Figure~S13}). All the electronic parameters, except the $\Delta_{LE-CT}$ energy gap, are derived from the electronic structure calculations and kept fixed across the solvents. The vibrational parameters are derived from the fits to the PDI monomer lineshape (\vt{Figure~S10}) and the observed vibrational quantum beats (\vt{Figure~S7}). }\\

 \section*{Data Availability}
The spectroscopic data generated in this study are available from the corresponding author upon a reasonable request.

\section*{Code Availability}
Python scripts for data processing, analysis and vibronic exciton simulations are available from the corresponding author upon a reasonable request.

\bibliography{SpPDI2}

\section*{Acknowledgements}
{\footnotesize S.P. and S.G. acknowledge the research fellowship from IISc Bangalore. A.B. acknowledges the Prime Minister's Research Fellowship, MoE, India. V.N.B. acknowledges the DST-Inspire research scholarship. J.S. acknowledges the University Grants Commission (UGC) for financial support. P.D.M thanks IISER TVM for financial assistance. M.H. acknowledges MoE-STARS/STARS-2/2023-0770 for financial support.}

\section*{Author Contributions}
{\footnotesize V.T. and M.H. conceptualized the project. S.P. optimized and performed the pump-probe and anisotropy experiments with help from V.N.B., A.B. and S.G. J.S. carried out the synthesis of SpPDI2 and performed the electronic structure calculations. P.D.M performed the synthesis and electronic structure calculations for the chiral CyPDI2 dimer. S.P. performed the vibronic exciton model calculations. V.T. wrote the manuscript with contributions from S.P. All authors discussed the results and commented on the manuscript. }

\section*{Competing Interests}
{\footnotesize The authors declare no competing interests.}

\end{document}